\documentclass[aps, prb, twocolumn]{revtex4}
\usepackage{amssymb}
\usepackage{graphicx}
\usepackage{hyperref}

\begin{document}

\title{Absence of structural correlations of magnetic defects in heavy fermion LiV$_\textbf{2}$O$_\textbf{4}$}
\author{S. Das}
\author{A. Kreyssig}
\author{S. Nandi}
\author{A. I. Goldman}
\author{D. C. Johnston}
\affiliation{Ames Laboratory and Department of Physics and Astronomy, Iowa State University, Ames, Iowa 50011}

\date{\today}
    
\begin{abstract}

Magnetic defects have pronounced effects on the magnetic properties of the face-centered cubic compound ${\rm LiV_2O_4}$. The magnetic defects arise from crystal defects present within the normal spinel structure. High-energy x-ray diffraction studies were performed on ${\rm LiV_2O_4}$ single crystals to search for superstructure peaks or any other evidence of periodicity in the arrangement of the crystal defects present in the lattice. Entire reciprocal lattice planes are mapped out with help of synchrotron radiation. No noticeable differences in the x-ray diffraction data between a crystal with high magnetic defect concentration and a crystal with low magnetic defect concentration have been found. This indicates the absence of any long-range periodicity or short-range correlations in the arrangements of the crystal/magnetic defects. 

\end{abstract}
\maketitle
\section{\label{intro}introduction}

${\rm LiV_2O_4}$ is a material of great interest as it shows heavy fermion behavior at low temperatures ($T \lesssim$ 10 K) in spite of being a \textit{d}-electron metal.\cite{Kondo1997} This is of particular interest because most of the well known heavy fermion compounds have crystallographically ordered arrays of \textit{f}-electron atoms. ${\rm LiV_2O_4}$ has a face-centred cubic crystal structure (space group $F d \overline{3} m$). Each V atom is coordinated with six O atoms to form a slightly distorted octahedron.\cite{Das2007} The V atoms themselves form corner sharing tetrahedra, often called the ``pyrochlore lattice'', which is strongly geometrically frustrated for antiferromagnetic ordering. The vanadium atoms with nominal oxidation state of $+3.5$ and occupying equivalent sites in the structure make ${\rm LiV_2O_4}$ metallic. The heavy fermion nature of ${\rm LiV_2O_4}$ was discovered to occur below $\sim$ 10 K from measurements of a large $T$-independent magnetic susceptibility $\chi$ $\sim 0.01$ cm$^3$/mol and a large Sommerfeld coefficient $\gamma \sim 420$ mJ/mol K$^2$\@.\cite{Kondo1997} 

Magnetic defects in the structure have a pronounced effect on the magnetic properties of ${\rm LiV_2O_4}$. For both polycrystalline samples and single crystals with extremely low magnetic defect concentration ($n_{\rm defect}$ $\leq$ 0.01~mol\%), the low-$T$ $\chi$ is $T$-independent.\cite{Kondo1997,Kondo1999,Das,Matsushita} The heavy fermion behavior of ${\rm LiV_2O_4}$ referred to above was inferred from measurements on such samples with extremely low $n_{\rm defect}$. However, in both powder and single crystal samples of ${\rm LiV_2O_4}$ with high $n_{\rm defect}$ (up to a maximum of 0.8 mol\%), the magnetic susceptibility shows a Curie-like upturn at low $T$\@.\cite{Kondo1997,Kondo1999,Das,Das2007} Crystal defects are the only possible source of these magnetic defects since magnetic impurity phases as the source of the low $T$ Curie-like upturn was ruled out.\cite{Kondo1999,Das} Low $T$ magnetization measurements on ${\rm LiV_2O_4}$ polycrystalline and single crystal samples containing magnetic defects revealed large values of the average spins of these defects.\cite{Das2007,Kondo1999,Das} The spin values $S_{\rm defect}$ range from $\sim$~2 to 4.

The presence of magnetic defects has a dramatic influence on $^7$Li NMR measurements of ${\rm LiV_2O_4}$. NMR measurements on polycrystalline samples of ${\rm LiV_2O_4}$ with extremely low $n_{\rm defect}$ show a linear variation of the $^7$Li spin-lattice nuclear relaxation rate ($1/T_1$) versus $T$ at low $T$\@.\cite{Johnston2005} This is typical for Fermi liquids. However, for polycrystal samples of ${\rm LiV_2O_4}$ with higher amounts of magnetic defects, the $^7$Li $1/T_1$ shows a peak at $\sim$ 1 K, and the relaxation recovery becomes strongly nonexponential.\cite{Johnston2005,Zong2008} This observation raises the question whether the ground state of a ${\rm LiV_2O_4}$ sample with high $n_{\rm defect}$ is still a Fermi liquid or is a non-Fermi liquid. If the ground state changes to a non-Fermi liquid, then there might be a critical $n_{\rm defect}$ for the transition.  Johnston et al.\cite{Johnston2005} suggested a model in which a crystal defect locally lifts the geometric frustration and thus allows magnetic order over a finite region around that defect, called a magnetic droplet. This model is qualitatively consistent with the large average values of $S_{\rm defect}$ obtained from the low $T$ magnetization measurements.\\

Given the pronounced effects of the magnetic defects on the properties of ${\rm LiV_2O_4}$, it is very important to examine if there are any periodic correlations in the distribution of the crystal defects which produce the magnetic defects or if they are randomly distributed. If we assume that a single crystal defect gives rise to a single magnetic defect, then the concentrations of crystal defects are extremely small ($<$~0.8 mol\%) to produce any observable change in intensities of x-ray Bragg reflections. One way to look for such small effects is to map out complete reciprocal planes and search for features in addition to the normal Bragg reflections. Any long-range periodicity of the crystal defects would produce additional peaks in the x-ray diffraction patterns, and short-range ordering would cause streaking of the Bragg peaks. Here we report on the high-energy x-ray studies of single crystals of ${\rm LiV_2O_4}$ with different magnetic defect concentrations.

\section{\label{expt}Experimental Details}

High quality single crystals of ${\rm LiV_2O_4}$ used in the experiment were grown in a vertical tube furnace using ${\rm Li_3VO_4}$ as the flux.\cite{Das2007} Three crystals, denoted as crystal 2, crystal 9, and crystal 10 with respective masses of 0.354 mg, 6.548 mg, and 2.1 mg were examined. The magnetic measurements on the crystals were done using a Quantum Design superconducting quantum interference device (SQUID) magnetometer in the temperature range 1.8~--~350 K and magnetic field range 0~--~5.5 T\@. The high-energy x-ray difraction measurements were performed at the 6-ID-D station in the MU-CAT sector of the Advanced Photon Source, Argonne National Laboratory. The energy of the radiation was set to 100 keV to ensure full penetration of the sample. The beam size was $0.3\times0.3$ mm$^2$. To record the full two-dimensional patterns, a MAR345 image-plate was positioned 705 mm behind the sample. During the experiments, the crystals were set between two pieces of thin kapton film and mounted on the sample holder.

\section{\label{magnetization}magnetic susceptibility and magnetization}

\begin{figure}
\includegraphics[width=3in]{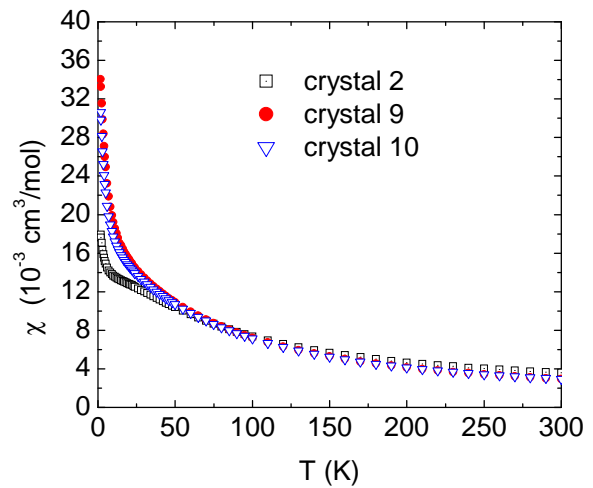}
\caption{Magnetic susceptibilities $\chi$ versus temperature $T$  of ${\rm LiV_2O_4}$ crystals containing different concentrations of magnetic defects. The susceptibilities are measured in 1 T magnetic field.}
\label{susc}
\end{figure}

\begin{figure*}
\includegraphics[width=5in]{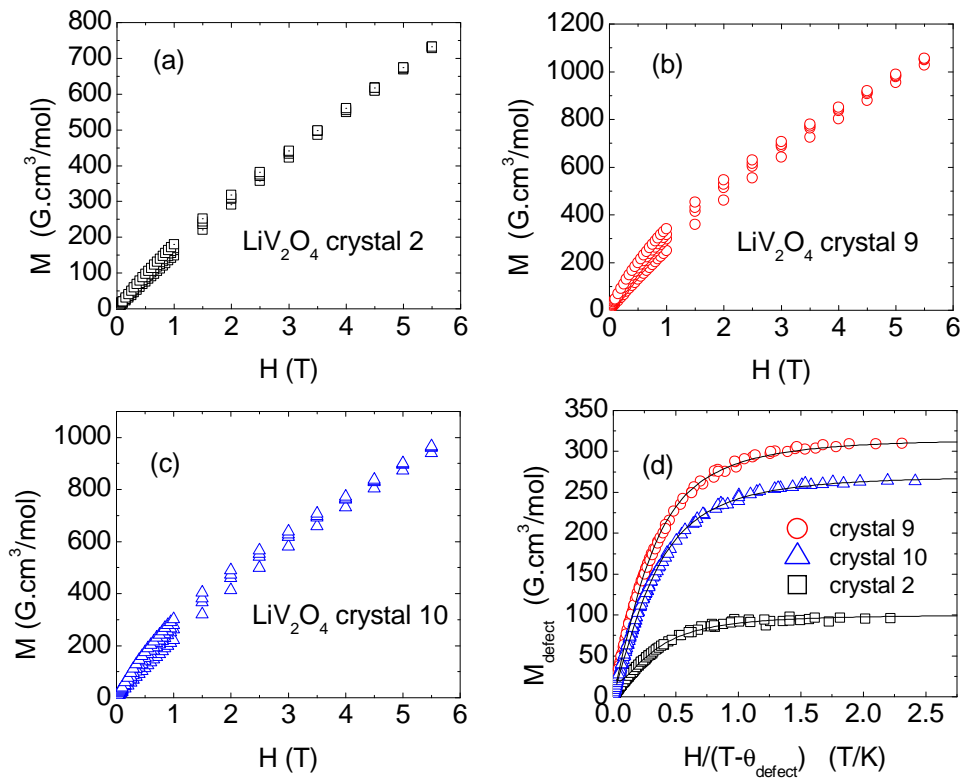}
\caption{Molar magnetization $M$ versus applied magnetic field $H$ isotherms at low temperatures ($T \leq 5$ K) for crystals (a) 2, (b) 9, and 10 (c), respectively. The four data sets shown in each of Figs.~\ref{allimp} (a), (b), and (c) are the $M(H)$ isotherms at four different temperatures 5 K, 3 K, 2.5 K, and 1.8 K\@. Figure~\ref{allimp}(d) shows the magnetic defect contribution to the magnetization for each crystal ${\rm \textit{M}_{defect}}$ = $M~-~\chi H$ plotted versus $H/(T - \theta_{\rm defect})$. The solid lines are the plots of Eq. (\ref{fiteq}) for each crystal with the parameters listed in Table~\ref{table1}\@.}
\label{allimp}
\end{figure*}

Figure~\ref{susc} shows the magnetic susceptibility $\chi$ versus temperature $T$ of the crystals 2, 9, and 10 measured in a 1 T magnetic field. The magnetic defect concentrations of the crystals were calculated by fitting the observed molar magnetization $M$ isotherms at low temperatures [$T\leq 5$ K, shown in Figs.~\ref{allimp}(a), (b), and (c)] by the equation\cite{Kondo1999,Das} 
\begin{equation}
M = \chi H + n_{\rm defect} N{\rm _A}g{\rm _{defect}}\mu{\rm _B} S{\rm _{defect}} B_{S}(x)\, , 
\label{fiteq}
\end{equation}
where ${n_{\rm defect}}$ is the concentration of the magnetic defects, $N{\rm _A}$ is Avogadro's number, $g{\rm _{defect}}$ is the $g$-factor which was fixed to 2 for the spins of the magnetic defects (the detailed reasoning behind this is given in Ref.\ [\onlinecite{Kondo1999}]), $\mu_{\rm B}$ is the Bohr magneton, $S{\rm _{defect}}$ is the average spin of the defects, $B _S(x)$ is the Brillouin function, $\chi$ is the intrinsic susceptibility of the pure ${\rm LiV_2O_4}$ spinel phase in the absence of magnetic defects, and $H$ is the applied magnetic field. The argument of the Brillouin function $B _S(x)$  is $x$~=~$g{\rm _{defect}}$$\mu{\rm _B}$$S{\rm _{defect}}$$H$/[k${\rm _B}$($T$$-$$\theta{\rm _{defect}}$)] where $\theta{\rm _{defect}}$ is the Weiss temperature associated with the magnetic defects and k${\rm _B}$ is Boltzmann's constant. The parameters fitted are $\chi$, $n_{\rm defect}$, $S{\rm _{defect}}$, and $\theta{\rm _{defect}}$.

The best-fit parameters obtained from the fits of the $M(H)$ isotherm data in Figs.~\ref{allimp}(a), (b), and (c) by Eq. (\ref{fiteq}) are tabulated in Table~\ref{table1} for each crystal. Figure~\ref{allimp}(d) shows the defect contributions to the magnetization ${\rm \textit{M}_{defect}}$ = $M - \chi H$ for each crystal plotted versus $H/(T - \theta_{\rm defect})$. All the data points in Figs.~\ref{allimp}(a),~(b),~and (c) collapse onto a universal curve for each crystal, thus verifying the validity of the model and the fits. The solid lines in Fig.~\ref{allimp}(d) are the plots of Eq. (\ref{fiteq}) for the three crystals with the parameters listed in Table~\ref{table1}, respectively. 

\begin{table*}
\caption{Magnetic parameters obtained from the magnetization versus field measurements below 5 K of the three ${\rm LiV_2O_4}$ crystals. $\chi$, $n{\rm _{defect}}$, $S$${\rm _{defect}}$, and $\theta{\rm _{defect}}$ are magnetic susceptibility, magnetic defect concentration, spin of the magnetic defects, and Weiss temperature of the interactions among the magnetic defects, respectively. The numbers in parentheses indicate errors in the last digit of a quantity.}
\begin{ruledtabular}
\begin{tabular}{llllll}
Sample no & $\chi$ (cm$^3$/mol) & $n{\rm _{defect}}$ (mol\%) & $S$${\rm _{defect}}$ & $\theta{\rm _{defect}}$ (K) & $n{\rm _{defect}}$$S$${\rm _{defect}}$ (mol\%)\\
\hline
crystal 2 & 0.01158(6) & 0.24(1) & 3.6(2) & $-$0.7(1) & 0.86(1)\\
crystal 9 & 0.0135(1) & 0.71(3) & 3.9(1) & $-$0.6(1) & 2.78(7)\\
crystal 10 & 0.0127(1) & 0.67(2) & 3.6(1) & $-$0.5(1) & 2.38(6)\\
\end{tabular}
\end{ruledtabular}
\label{table1}
\end{table*}

\section{\label{xray}High-energy x-ray diffraction measurement}

In order to search for long-range or short-range order in the arrangement of the crystal defects giving rise to the magnetic defects within the crystal structure, we carried out high-energy synchrotron x-ray diffraction measurements over a wide range of reciprocal space. The rocking technique used to record the diffraction intensities from planes in reciprocal space has been described in detail in Ref. [\onlinecite{kreyssig}]. The patterns were obtained by recording the Bragg reflections of all points of a reciprocal lattice plane intersecting the Ewald sphere. The orientation of the reciprocal lattice relative to the Ewald sphere is given by the orientation of the crystal. So, tilting the crystal turns complete reciprocal lattice planes of the crystal to intersect the Ewald sphere, depending on the range of tilting angle. In the experiment, diffraction patterns were obtained by tilting the crystal through two independent angles $\mu$ and $\eta$ perpendicular to the incoming x-ray beam. Patterns were recorded by continously scanning through $\mu$ as $\eta$ was increased in small steps. By averaging the recorded patterns obtained at different values of $\mu$ and $\eta$, a considerable range of the designated reciprocal lattice planes was mapped out. This averaging over large parts of a Brillouin zone also enhances very weak broad scattering features making them detectable.

Depending on the kind of modification/deviation of the crystal structure arising from the crystal defects, we expect to see different modifications/deviations in the diffraction patterns of the reciprocal planes. A crystallographic superstructure, either commensurate or incommensurate, will produce weak additional Bragg reflections. Lower-dimensional or short-range order will produce broad features or diffuse scattering. For example, a two-dimensional order yields a rod-like scattering feature. If the incoming beam is parallel to the axis of the rod, we will see a spot in the diffraction pattern of that plane. The same feature, however, will yield a streak of intensity in the diffraction patterns of reciprocal planes perpendicular to the rod.

\begin{figure*}
\includegraphics[width=4.8in]{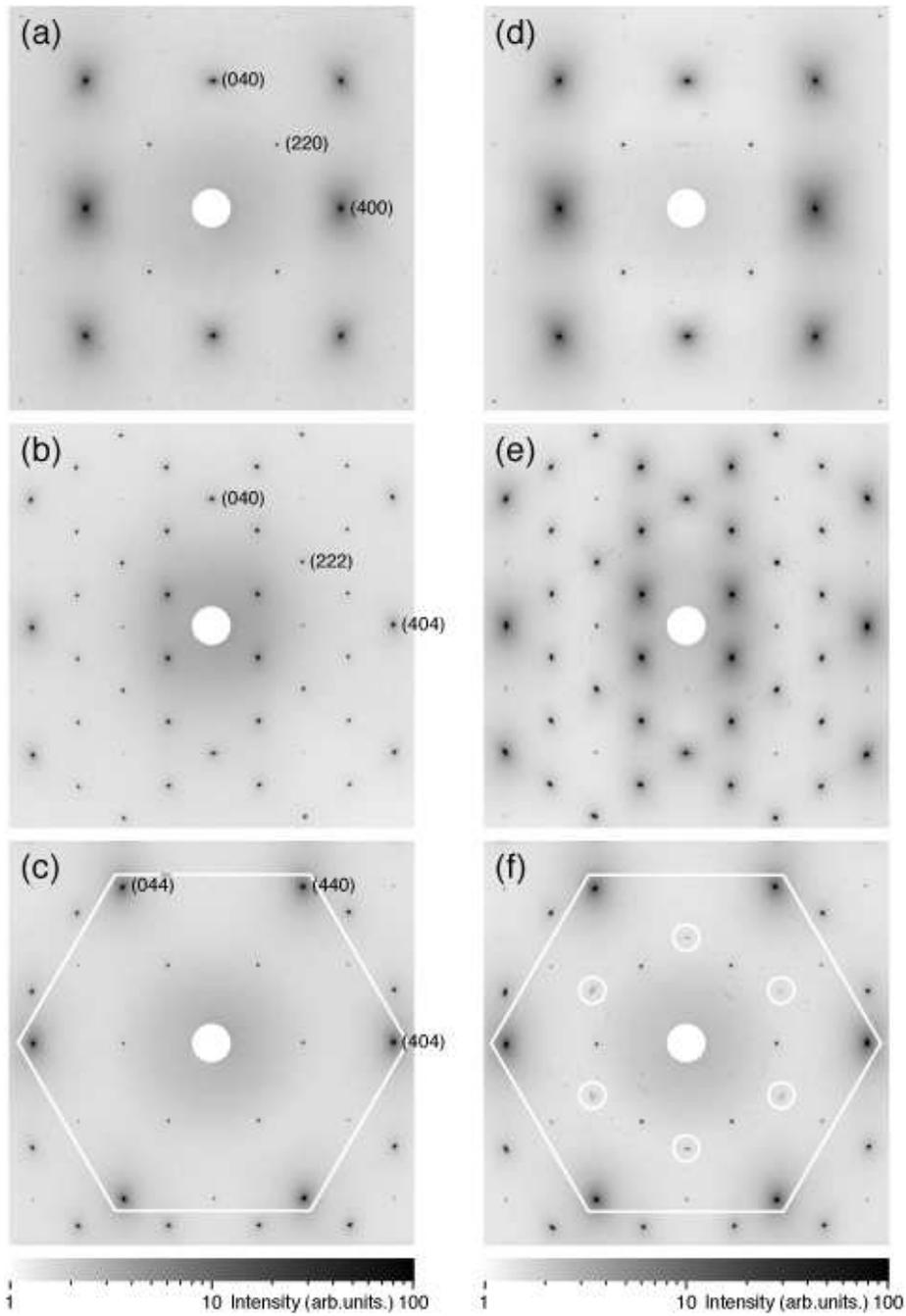}
\caption{High-Energy x-ray diffraction patterns of ${\rm LiV_2O_4}$ single crystals 2 and 9. Figures~\ref{both}(a), (b), and (c) show the patterns for reciprocal planes of crystal 2 perpendicular to the (001), ($\overline{1}$01), and ($\overline{1}$11) directions, respectively. Figures~\ref{both}(d), (e), and (f) show the patterns for reciprocal planes of crystal 9 perpendicular to the (001), ($\overline{1}$01), and ($\overline{1}$11) directions, respectively. In (f), the extended features indicated by white circles are reflections that are not allowed by the symmetry. The reflection spots enclosed by the white lines in (c) and (f) are from one Brillouin zone. The spots outside the white polygon are from other Brillouin zones in the same orientation.}
\label{both}
\end{figure*}

In our experiment, reflections from reciprocal lattice planes perpendicular to the three high symmetry directions, namely (001), ($\overline{1}$01), and ($\overline{1}$11) of the cubic structure, were recorded. Figures~\ref{both}(a), (b), and (c) show the room temperature diffraction patterns from planes in the reciprocal space of crystal 2 (${\rm \textit{n}_{defect}}$ = 0.24 mol\%) perpendicular to (001), ($\overline{1}$01), and ($\overline{1}$11) directions, respectively. The lattice planes perpendicular to the ($\overline{1}$11) direction are very closely spaced. Thus in this direction, when we tilt the crystal, higher order reciprocal planes will also intersect the Ewald sphere and be observed.\cite{kreyssig} This was minimized by limiting the range of tilting. In Fig.~\ref{both}(c), only those spots enclosed by the white lines are due to Bragg-reflections from the Brillouin zone containing the origin. The spots outside the polygon arise from other Brillouin zones. 

In Figs.~\ref{both}(a), (b), and (c), all the spots observed are allowed by the space group of the crystal. The intensity at the center of the Bragg reflections are 3 -- 6 orders of magnitude higher than counts shown in black at the maximum in the scale for the contour map. We used iron slabs up to 3 cm in thickness to increase the dynamic range from $10^4$ (intrinsic for the detector) to $10^7$ by attenuating the incident x-ray beam. No extra spots or diffuse scattering were observed in the patterns. The shape of the spots is also as expected for the given resolution conditions. Thus we conclude that crystal 2 is almost a perfect crystal. There are no other single crystals or grains oriented in other directions. The small areas, a little darker than the background and surrounding the sharp black Bragg reflection spots, are the tails of the Bragg reflections.  

Figures~\ref{both}(d), (e), and (f) show the room temperature x-ray diffraction patterns from reciprocal planes of crystal 9 (${\rm \textit{n}_{defect}}$ = 0.62 mol\%) perpendicular to (001), ($\overline{1}$01), and ($\overline{1}$11) directions, respectively. For the planes perpendicular to the (001) and ($\overline{1}$01) directions, there are no differences between the patterns obtained for crystal 2 and crystal 9. For the plane perpendicular to the ($\overline{1}$11) direction, a few spots were observed marked by solid circles in Fig.~\ref{both}(f), which are not allowed by the symmetry of the space group and are missing in Fig.~\ref{both}(c). These extra features are extended and have intensity $10^{-5}$ times that of the Bragg reflections. Twinning or stacking faults of similiarly oriented crystals can cause such features to appear. But as seen from the spots outside the white polygon, the features are not present in the other Brillouin zones with the same orientation. This excludes the possibility of periodic arrangement of such crystal defects. The slight broadening of the peaks in Fig.~\ref{both}(e) is also due to the stacking faults or to a slight misalignment of the crystal.

\begin{figure}
\includegraphics[width=3in]{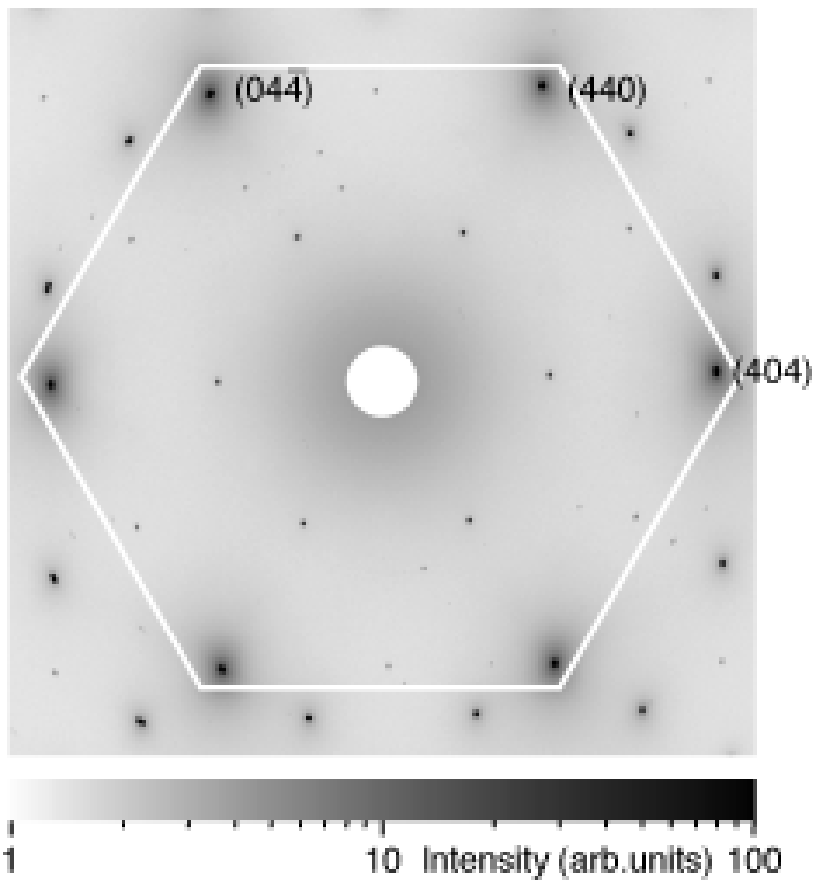}
\caption{High-energy x-ray diffraction pattern of the reciprocal lattice plane perpendicular to the ($\overline{1}$11) direction of ${\rm LiV_2O_4}$ crystal 10.}
\label{other}
\end{figure}

To test if the appearance of the extended extra features for crystal 9 is an artifact of the particular crystal or is intrinsic, we performed the same experiment on crystal 10 which was grown under similiar conditions and has a similiar magnetic defect concentration as that of crystal 9. The x-ray diffraction pattern for the reciprocal lattice plane perpendicular to the ($\overline{1}$11) direction of crystal 10 is shown in Fig.~\ref{other}. The extra extended spots present in Fig.~\ref{both}(c) are missing here. There are a few very weak spots other than the allowed ones for the plane perpendicular to ($\overline{1}$11) direction. These are caused by other misaligned crystals of the same material or impurities and show the very high sensitivity of the technique to the smallest deviations/differences in the pattern from the expected one for a perfect crystal. If we focus our beam onto a different spot on the same crystal surface, the extra peaks vary in intensity and/or disappear. 

\section{\label{summary}Summary}

No noticeable difference in the x-ray diffraction patterns of the reciprocal lattice planes of a crystal with high magnetic defect concentration and a crystal with low magnetic defect concentration has been found. This indicates the absence of any long-range periodicity or order in the arrangement of the crystal defects giving rise to the magnetic defects. We also did not observe any rod-like features or extra planes in reciprocal space and thus exclude any long-range low-dimensional order. No diffuse scattering was observed in the dynamic range of five orders of magnitude. This excludes the possibility of short-range order of the crystal defects. Thus we conclude that the crystal defects in ${\rm LiV_2O_4}$ which produce the magnetic defects are randomly distributed within the spinel structure.

\begin{acknowledgments}
We thank D. Robinson for assistance with the x-ray diffraction measurements at the Advanced Photon Source. Work at the Ames Laboratory was supported by the U.S. Department of Energy, Basic Energy Sciences under Contract No. DE-AC02-07CH11358. Use of the Advanced Photon Source (APS) was supported by the U.S. Department of Energy, Basic Energy Sciences, under Contract No. DE-AC02-06CH11357. The Midwest Universities Collaborative Access Team (MUCAT) sector at the APS is supported by the U.S. Department of Energy, Basic Energy Sciences, through the Ames Laboratory under Contract No. DE-AC02-07CH11358.
\end{acknowledgments}

\end{document}